# The evolvability of business and the role of antitrust

## BY IAN WILKINSON*

### Discipline of Marketing, The University of Sydney, Australia

I     Introduction

Complexity science is relevant to Antitrust Policy because the economy is a complex adaptive system of economic and non-economic actors.[1] The study of complexity focuses our attention on the way complex behaviour in natural, biological, social or economic systems arises from interactions among the parts rather than from any inherent complexity of the parts; simple rules of interaction can give rise to complex system behaviour over time. The importance of feedback effects and the way systems, learn, adapt and evolve are also emphasised, which are the processes by which the rules of interaction and mix of players in a system change over time. As Stuart Kauffman nicely expresses it – the winning games are the games the winners play.[2]

In this paper, based on theories of complex adaptive systems, I argue that the main case for antitrust policy should be extended to include the criteria of "evolvability."[3] To date, the main case focuses on economizing, including market power as a key filter for identifying suspect cases.[4] Both production and transaction costs are considered as part of economizing and other factors are use to consider the benefits of different industry structures. CAS analysis focuses attention on dynamics, evolution and networks. As I will show, the criteria of evolvability requires us to consider various types of direct and indirect network impacts in business that go beyond the traditional focus on production and transaction costs. These network impacts stem from the connections between transactions and relations over time and place, including how business arrangements at one time, limit or enable arrangements in the future. An assessment of the impacts, I argue, can and should be included in the rules of antitrust and in the processes of antitrust case analysis and decision making.

---

[1] W. Brian Arthur, Steven N. Durlauf and David A. Lane eds. THE ECONOMY AS AN EVOLVING COMPLEX SYSTEM II (1997).

[2] Stuart Kauffman INVESTIGATIONS (2000).

[3] I use the term "main case" in the same way as Oliver E. Williamson does in his book THE MECHANISMS OF GOVERNMENT (1996).

[4] Ibid.

Much of what I have to say is not really new to economic and business theory. My ideas about complexity and business build on the theories of institutional evolution of Douglas North,[5] the pioneering work of Nelson and Winter on evolutionary econmics,[6] the concerns of cutting edge economists[7] and the work of researchers on economic and business systems associated with the Santa Fe Institute and Agent Based Computational Economics.[8] But I believe the time is ripe now to bring these ideas together and to see what they mean for public policy and, in particular, antitrust policy.

Developments in the so called "complexity sciences" - the study or complex adaptive systems (CAS) – provide ways for us to understand, analyse, map and model business network impacts over time and place that can assist antitrust decisions and policy development. Of particular importance are new types of methodologies for modelling the dynamics and evolution of CAS using agent based models. These models are not the same as the statistical and mathematical methods we use to in analysing firm and industry behaviour. Traditional methods cannot deal with the inherent nonlinearity of CAS, which arise from interaction and feedback effects taking place among the parts, and the consequent multiple equilibria and behaviour regimes that arise. The only way to study such systems is to play out their behavior over time via computer simulations using agent based models. Such simulations play the same role as mathematical and statistical methods; they force us to be explicit about the rules governing behaviour and interactions and about contextual assumptions. The logic of the program reveals the implications of the rules and assumptions for system behaviour. It is from simulation studies of this type, as well as other types of numerical simulations, that the hallmarks of CAS behaviour have been revealed, including sensitivity of outcomes to starting conditions, path dependence, bifurcations or tipping points, basins of attraction and the stability of different system attractors. The term Artificial Life is sometimes used to characterize these types of models because they mimic the processes of living, reproducing and evolving. As Chris Langton, one of the founding fathers, observed, we are restricted to the study of what nature has left around for us to study and this results from one play of the tape of life – life as it is.[9] Artificial Life methods allow us to consider alternative evolutionary conditions and pathways - to investigate life as it could be. As I will show, these types of models and simulations offer a way forward for Antitrust case analysis and decision-making.

---

[5] Douglas North Economic Performance through Time NOBEL PRIZE LECTURE (1993).

[6] Richard R. Nelson and Sidney Winter AN EVOLUTIONARY THEORY OF ECONOMIC CHANGE (1982).

[7] David Colander, Richard P.F. Holt, and J.B. Rosser Jr. THE CHANGING FACE OF ECONOMICS (2004).

[8] see for example Arthur et al supra (1997).

*A    Links to Antitrust*

Antitrust policy is about influencing the rules of interaction and evolution of business systems, with the aim of avoiding pathological evolutionary paths and the emergence, survival and reproduction of undesirable firms and business systems. My reading of recent critiques of antitrust policy indicate that current policies focus primarily on problems of static market efficiency and price competition rather than dynamic market efficiency and value. Static market efficiency is reflected in rules to preserve price competitive markets and to limit market power, which is interpreted in terms of the distribution of market shares in horizontally defined domestic markets for substitutable and similar products or services. Dynamic market efficiency has to do with the development and evolution of new types of markets, firms and industries that create and deliver value to consumers. While there are links between static and dynamic market efficiency these are problematic. For example, larger, more powerful, firms may be able to devote more resources to innovation and thereby aid the evolutionary process. But they may also block or suppress undesirable (from their point of view) new competitors and be unable to recognize new types of opportunities that potentially undermine their existing position.

---

[9] Chris Langton ed. ARTIFICIAL LIFE (1996).

Michael Porter is one who has argued for antitrust policy to focus on value, rather than just price and cost, and on dynamic rather than static efficiency, and he advocates the use of his frameworks for analyzing competition in industries and nations to assess antitrust cases.[10] His research also suggests that variation in market shares rather than their mean distribution is a more effective indicator of dynamic efficiency.[11] This is as a step forward in our understanding and measurement of dynamic efficiency in business systems but a clear underlying theory of business system dynamics and evolution is not articulated. Instead, a list of factors to consider in each case is presented, which hint at the underlying processes to which antitrust policy needs to direct its attention. I believe that ideas from complexity theory and theories of cultural evolution can help enrich and advance our understanding of the key processes driving the evolution of firms, markets and industries and provide a more coherent focus for antitrust policy.

---

[10] Michael M. Porter Competition and Antitrust: A Productivity-Based Approach to Evaluating Mergers and Joint Ventures ANTITRUST BULL 46 (2001).

[11] M. Sakakibara, M. and Michael M. Porter Competing at Home to Win Abroad: Evidence from Japanese Industry THE REVIEW OF ECONOMICS AND STATISTICS 83 (2001).

The title of this article is inspired by one by Richard Dawkins, the eminent Neo-Darwinian biologist, entitled "The evolution of evolvability," which was presented at the first Artificial Life conference at the Santa Fe Institute in 1987.[12] The topic of evolvability goes to the heart of the problems confronting antitrust policymakers, who are concerned to influence, through regulation and enforcement, the process of evolution of business systems towards more desirable community outcomes, including efficiency and effectiveness. Evolvability is the essence of dynamic efficiency, whereas static efficiency is about tinkering with an existing business system to achieve improved outcomes, not with how this will affect the process of future evolution of the system.

The problem is that we cannot predict the future of business evolution, any more than we can predict the future of natural evolution. The history and existing nature of business systems affect how they can and cannot evolve but in ways we cannot fathom in advance because of the complex, nonlinear, self-organizing, adaptive nature of the systems involved. Complexity and evolutionary theory teach us how starting conditions and history matter and how micro interactions and adaptations lead to emergent macro patterns, and that systems gravitate towards different types of attractors depending on chance factors and tipping points that are only knowable from hindsight.

At first sight this seems to be an argument for anything goes and for no antitrust policy because anyone's guess about the future is as good as anyone else's. I disagree. The role of antitrust policy is to help encourage and select business systems for their evolvability, not to control or predict evolution. It is crucial to realize and emphasize that antitrust policy makers and enforcers do not stand outside of the world of business, observing it in some godlike manner, issuing edicts and making up rules and enforcing them. It is part of the system. It affects, through its actions and responses, what business does and how it evolves. It is itself directly and indirectly affected by business, society and academia and, over time, it develops and evolves. Interactions with business and academia take place all the time, through lobbying activities, informal and formal meetings, confrontations in the courts and in its responses to what business and academics do and say. In short, a co-evolutionary process is going on involving antitrust policy makers and enforcers (including lawyers and Law Schools), business (including managers, workers, consumers, and other stakeholders) and academia (including economists and Business Schools). The problem for antitrust policy is not how to control and direct business in beneficial ways but to participate in the system of business in a way that contributes to the productive evolution of the evolvability of business systems.

---

[12] Richard Dawkins The Evolution of Evolvability in Langton *supra* (1987).

Business systems are living systems, business ecosystems,[13] and if they cannot evolve they will eventually stagnate and die, because they will be unable to cope with and contribute to changing business contexts, including new types of ideas, technologies, demands, competitors and natural conditions. In order to understand this more clearly, and what it means for antitrust, we need to understand the nature of evolutionary processes, starting with biology and then moving on to cultural systems, which includes business systems.

II      The Nature of the Evolutionary Process

Figure 1 depicts the main elements of evolution in terms of two key processes. First is the existence of entities that are capable of being reproduced over time. In biology the entities are the genes that are replicated as genotypes are passed from generation to generation. Genes are essentially subroutines that become expressed and used in some kind of order over time as a plant or animal develops into its phenotype.[14] In cultural evolution, including business culture, the entities that are replicated through time are called by different researchers' cultural traits, routines or competencies.[15] These are acquired or modified by social learning, including teaching, imitation and other forms of social transmission and affect behaviour.[16] Dawkins calls to the basic entities of cultural evolution *memes* that leap from mind to mind and are thereby reproduced, altered, reconfigured, and diffused through time and space.[17] Memes include ideas, knowledge, beliefs, values, skills, capacities, attitudes and orientations. But an individual idea is not really equivalent to a subroutine, as genes are, unless we define memes to include a related set of ideas that constitute a way of doing something. These could be called meme complexes but to keep things simple I will use the term meme to refer to the subroutines expressed in cultural and business life.

---

[13] James Moore THE DEATH OF COMPETITION (1996).

**\*\*\* Insert Figure 1 about here**

Genes and memes do not exist in isolation but form genotypes or memotypes i.e. assortments of cultural traits that characterize a particular person, group or firm, that govern the way a phenotype develops and behaves. In biology, phenotypes are the myriad of types of flora and fauna that develop from the population of genotypes existing in a generation. The success of phenotypes affects whether or not the genes governing their behaviour will be passed on to the next generation, whether they will survive. In cultural evolution, phenotypes refer to the characteristic patterns of behaviour and responses of people and organisations that are operating under the influence of different sets of memes or cultural traits. The term business model is sometimes used to refer to the mix of traits characterizing a particular type of firm's manner of operation and response and is a kind of business analogue to the concept of a genotype in biology. Just as in biology, the success of a business model or memotype in its environment affects which memes or traits get reproduced over time and place.

---

[14] Richard Dawkins THE DEVILS CHAPLAIN (2003).

[15] Richard R. Nelson and Sidney Winter AN EVOLUTIONARY THEORY OF ECONOMIC CHANGE (1982).

[16] Peter J. Richerson and Robert Boyd NOT BY GENES ALONE : HOW CULTURE TRANSFORMED HUMAN EVOLUTION (2004).

[17] Susan Blackmore *The Meme Machine* (1999).

Genes and memes do two things.[18] They influence the development of the phenotype and they get themselves reproduced. The success of different phenotypes depends on the environment in which they develop and operate. The environment comprises the material world, the world of physics, chemistry and biochemistry, as well as other phenotypes that co-exist and with which a phenotype interacts. Natural selection refers to the struggle among phenotypes to develop and survive in a particular environment, independently or in cooperation. The same principle applies to biological and cultural evolution.

For genes to get themselves copied into the next generation, more than survival is required. Sexual selection, Darwin's other theory as Dawkins refers to it,[19] focuses on the struggle among males and females to find and secure mates with whom they can cooperate in passing on their genes to future generations. Genes that are useful for sexual selection may not be the same as and may even conflict with genes for natural selection, leading to some weird and wonderful sexual dynamics. An analog to sexual selection exists in cultural and business evolution, which we may term "business mating."[20] This refers to the ability of firms to cooperate with other firms in achieving their goals.

---

[18] Dawkins *supra* (1987).

[19] Dawkins supra (2003).

The nature of business is commonly depicted in terms of natural selection with a focus on competition as the driving force. Firms continually struggle to survive in dynamic markets through creating and sustaining differential advantage.[21] But cooperation also matters and in recent years there has been an explosion of interest in the nature, role and importance of business relations and networks in shaping business behaviour and performance. The ability to form successful business relations with suppliers, customers, complementors and even competitors, and the ability to join and help co-produce mutually productive networks of such relations, requires competences that are different from but complementary to those involved in the struggle for competitive advantage. Firms cooperate to compete and compete to cooperate.[22] Firms are selected for their mating abilities as well as their competitive abilities and this creates selection pressures and transmission biases for particular types of business subroutines or competencies that could not be explained by natural or competitive selection and a focus on the firm as an isolated economic actor. Just as in the animal kingdom, people and firms struggle to choose and get chosen by better business mates and who mates with whom in business is not random, as assumed in perfectly competitive markets, but is a form of assortative mating. However, to date there has been only limited research directly addressing the issue of who mates with whom in business.[23]

---

[20] Ian F. Wilkinson, Louise C. Young and Per Freytag Business Mating: Who Chooses Whom and Gets Chosen? INDUSTRIAL MARKETING MANAGEMENT (2005).

[21] Wroe Alderson MARKETING BEHAVIOR AND EXECUTIVE ACTION (1957).

[22] Ian F. Wilkinson and Louise C. Young On Cooperating: Firms, Relations and Networks J. OF BUSINESS RESEARCH 55 (2002).

[23] Wilkinson et al *supra* (2005).

Table 1 summarizes the previous discussion and shows the terms used to describe key components of the evolutionary process in biology and business

**\*\*\* insert Table 1 about here**

I now focus on cultural evolution as this lies at the heart of our understanding of the development and spread of business systems and practices and the development of appropriate antitrust policies. The main processes that cause cultural change and evolution, that affect the number and mix of cultural variants in a focal population, have been summarized by Richerson and Boyd.[24] Inertial forces tend to reproduce the same cultural variants over time and result from unbiased sampling and faithful copying of memes. Forces for change comprise two types: (a) *transmission biases* that make people and firms more likely to encounter and adopt some memes rather than others and adapt their behaviour accordingly; and (b) *natural selection,* which affects what happens to people and firms that have different cultural variants or memotypes, whether they succeed or not and hence their memes are perpetuated or not, and whether they become models to copy or not. These processes represent generic types of targets for antitrust policy, as is explained in the next section.

B      *Normative Implications of an Evolutionary Perspective*

The evolutionary view has implications for firm strategy and for policy. For firms the strategic issues are how to effectively participate in the complex, self organizing, adaptive, evolving business systems of which they are necessarily a part. For antitrust policy the challenge is to influence transmission biases and natural selection processes so as to improve the evolvability of a nation's business systems of which it is a part, including: (a) weakening inertial forces that tend to reproduce the same business systems over time, when environmental variation calls for new business models;

---

[24] Richerson and Boyd *supra* (2004).

(b) shaping transmission biases to enhance productive entrepreneurial, innovation and imitation processes and (c) influencing natural selection processes, including the birth and death of firms, to ensure that the pool of cultural variants in the business population remains viable and varied, such that it opens up rather than narrows future development opportunities and evolutionary paths.

In order to understand the meaning of these type of policies we have to understand some of the main types of characteristics of CAS and the way they evolve.[25]

III       Main Characteristics of Complex Adaptive Systems

*A       History Matters*

History focuses attention on temporal and sequence effects, which are manifest in many ways. First, *starting conditions* affect the way a CAS behaves over time and the kinds of attractors that can emerge. This means that firms and business systems are restricted in what they can sense and the evolutionary pathways they can follow based on where they start from. Second, once we start down one evolutionary pathway it tends to block off some others, or what we refer to as *path dependence*. What is new about this? We always knew that firms could only act based on their resources, competences and orientations. Complexity teaches us a more profound lesson that the way we search forward in time and adapt is biased and may miss viable and better alternative evolutionary pathways and opportunities. We see this in research on innovation and the entrepreneurial process, which shows how prior knowledge affects our ability to recognize and discover new opportunities because it limits the kinds of links among existing and new ideas we are able to make.[26] Thus, collectively and individually we can only see and reach some futures. As the Irish joke goes, if you want to get to there I would not start from here! There is no guarantee that the assortments of knowledge that exist in the minds and memories of people and firms will allow us to recognize, let alone successfully exploit, all the opportunities available or even the best ones, and the ones that can recognize opportunities may not best able to exploit them.

---

[25] For further discussion see Ian F. Wilkinson and Louise C. Young Toward A Normative Theory of Normative Marketing Theory" J. OF MARKETING THEORY (2005).

The third way history matters is in terms of the *temporal order*, or sequence of events, and the tempo or rhythm of events. Simulations of complex systems show that order effects can change evolutionary paths and outcomes, not just the mix of types of events or factors influencing a system. History plays out over time and key events or sequences of events and factors, such as the entry order of competitors, the sequences of technologies, or confronting particular types of problems or selective disadvantages early rather than later in a firm and industry's development can have major impacts on subsequent patterns of success and development.[27] Earlier events set the stage, alter the starting conditions, for sensing and responding to later events, which in turn affect the kinds of opportunities that can and are recognized and acted on by firms. A particular example is that of bifurcation or tipping points, in which small apparently insignificant events can entrain patterns of development and evolution far into the future – the proverbial butterfly affecting global weather or the positive feedbacks resulting from one technology getting a head start e.g. Beta vs VCR.[28] Order effects are reflected in the learning and knowledge development processes in firms and industries, which enable and constrain future developmental pathways, such as of the kind identified by Levinthal and March,[29] Simon[30] and in Levitt's concept of marketing myopia.[31]

---

[26] Israel Kirzner Entrepreneurial Discovery and the Competitive Market Process: An Austrian Approach" J. OF ECON LIT XXXV (1997).

[27] Michael M. Porter THE COMPETITIVE ADVANTAGE OF NATIONS (1990).

[28] W. Brian Arthur Complexity and the Economy SCIENCE 284 (1999).

The effects of history are hard to see when much empirical analysis in business, economics and social science is cross sectional and variance based.[32] One of the lessons of research on complexity is to highlight the impact of historical processes and contingencies. Moreover, the increasing pace of change, speed of communication and far reaching inter-dependencies among businesses and economic systems across and within industries, technologies and nations makes their impact more difficult to ignore.

What can be done? At heart the task is to increase the ability to make productive links among ideas, both new and old, as well as facilitating the exploitation of the opportunities so recognized. There is a principle that seems relevant first espoused by Wroe Alderson, one of the founders of modern marketing theory. He called it the *power principle*: to act so as to promote the ability to act.[33]

Much research has focused on the way new ideas and innovations arise and diffuse in a social or business system. Much of it falls under the heading of entrepreneurship and it has become a hot topic and course offering in many business schools of late. The focus of research in this area used to be on the characteristics of the entrepreneur rather than the opportunity discovery process itself, but this has changed. The focus has shifted to examining the way different types of opportunities are discovered or discoverable. A critical role is played by the prior knowledge people and firms have, their access to new types of information, and the way new ideas come from combining existing ideas in new ways.[34]

---

[29] David A. Levinthal and James A. March The Myopia of Learning STRATEGIC MANAGEMENT J. (1993).

[30] Herbert Simon, H. Strategy and Organizational Evolution STRATEGIC MANAGEMENT J. 14 (1993).

[31] Theodore T. Levitt THE MARKETING IMAGINATION (1986).

[32] Gary J. Buttriss and Ian F. Wilkinson From variables to event based models of business J OF INTERNATIONAL ENTREPRENUERSHIP (in press).

[33] Alderson *supra* (1957).

One way of priming the innovation process is to facilitate new types of potentially productive knowledge combinations. For example, I am involved in research trying to develop intelligent software agents to mine the internet for knowledge combinations that could prime the opportunity discovery process.[35] Another development in Australia is the Bridge network concept base on the ideas (memes) of John Wolpert.[36] This initiative links firms in high tech industries in networks via trusted intermediaries who overcome some of the barriers to proprietary information sharing. In this way some of the potential opportunities from combining knowledge across firms in the network are identified that otherwise would not be seen. The potential opportunities are a function of the assortment of firms and knowledge that exist within the network and the potential productivity of new knowledge combinations from within that assortment.

Restrictions on the flow of information that limit the kinds of new knowledge combinations that can arise necessarily constrain the innovation and evolutionary process. Most of the great inventions of the past such as language, organisations, the printing press, intermediaries, the transistor, computers, high speed travel, communication and the internet have their impact not so much directly but from the way they free up the flow of information and allow productive new knowledge combinations (opportunities) to be discovered and acted on. See for example David Bodanis's summary of the myriad consequences of the transistor.[37] Each of these significant inventions represent a new platform for evolution to work on and are the cultural equivalent of the "invention" in nature of replicators, multi-cellular organisms, life, communication systems, mammals and minds.[38]

---

[34] Scott Shane Prior knowledge and the discovery of entrepreneurial opportunities ORGANIZATION SCIENCE 11 (2000).

[35] For more details see http://research.it.uts.edu.au/emarkets/.

[36] John Wolpert Breaking out of the Innovation Box HARVARD BUSINESS REVIEW August (2002).

Restrictions on the formation of potentially productive knowledge links can take many forms and antitrust law may wish to consider some of these. For example, one way new knowledge combinations arise is when people move between firms. But contractual clauses that limit disclosure and the types of firms an ex-employee can work for obviously limit the kinds of knowledge combinations that can occur. Against this is the issue of protecting the rights of those who have valuable knowledge to exploit and the incentives to find and exploit new opportunities.

On a more general level, the ability of a population of organisations to evolve and adapt in superior ways to each other and to changing environmental conditions depends on the variety within and between the organisations. This idea goes back to Ross Ashby's original concept of *requisite variety*, which states that the ability of a system to deal with its environment depends on matching the variety of the environment within the responding system.[39] In business systems the variety is reflected in the assortment of cultural variants or memotypes in the population of firms. Research on biological systems shows that the same or similar phenotypes may arise from different genotypes and co-exist in the population. When environmental conditions change, this variety in the genotypes plays an important role in providing different pathways forward. Genotypes will vary in how they respond and survive in the changed conditions and hence explore a wider space of possibilities. But if all the genotypes were the same, for example optimized to the current conditions, this would limit the response alternatives to random mutations and local differences in context.

---

[37] David Bodanis ELECTRIC UNIVERSE (2005).

[38] Dawkins *supra* (1989).

[39] W. Ross Ashby DESIGN FOR A BRAIN (1952).

Similar arguments apply to the evolvability of business systems and the range of generating memotypes underlying current firm behaviour. We know from studies of business evolution that radically new types of products and industries often come from outside the existing mainstream, in part because the mainstream tends to become myopic and restricted in their ability and willingness to recognize and exploit new types of business models, especially those that undermine traditional ways of thinking and behaving, so called disruptive technologies.[40]

B    *Interaction and Feedbacks Matter*

The second major contribution of complexity theory is to focus attention on the connections between the parts of a system as opposed to the properties of the individual parts. There are three aspects to this: the way micro interactions taking place among existing parts drives overall system behavior, the way overall system behavior has feedback effects on the behavior of the parts, and the way the pattern of interactions over time creates the parts themselves.

A complex adaptive system comprises a *network* of interconnected, interacting entities, actors or agents and overall system behaviour arises from the micro interactions taking place among them in a bottom up fashion. In addition the overall or macro behavior of the system has feedback effects on the parts, such as the way moves in share price indicators affect individual actors tock market behavior, in a top down fashion. This perspective contrasts with much of business and economic theory, which focuses on the number and properties of individual economic actors considered in isolation, including their decision-making and management characteristics.[41] In economic and business systems interactions and feedback effects among activities, actors, resources and ideas or schemas all play an important role in shaping behaviour and evolution.[42] A system cannot be reduced to the behavior of its parts, interactions matter. In business systems, these interactions are the means by which resources are accessed and created, problems are and opportunities are identified and confronted, innovations and adoptions occur, knowledge is shared and developed and value is created and developed.[43]

---

[40] Clayton Christensen *The Innovators Dilemma* (2000).

The implication for antitrust policy is that the relevant units of adaptation and evolution are not individual firms competing in a focal horizontal market but networks of interconnected, interdependent, interacting firms and other organisations spanning various markets, industries and technologies. These networks together create and deliver value to intermediate and end consumers and develop and co-evolve over time through their internal and external actions, responses and interactions. This involves a continual process of configuring and reconfiguring the connections between actors in the network, changes in the actors in the network and the role they play and the creation and destruction of new types of actors and relations. This co-evolutionary process cannot be understood, the main forces that drive development and evolution cannot be identified, and the potential role and impact of antitrust policy cannot be fathomed without focusing on the structure and behaviour of relevant business networks.

---

[41] See for example Phillip Ball CRITICAL MASS: HOW ONE THING LEADS TO ANOTHER (2005).

[42] Catherine Welch and Ian F. Wilkinson Idea Logics and Network Theory in Business Marketing J. OF BUSINESS TO BUSINESS MARKETING 8 (2002).

[43] See Wilkinson and Young *supra* (2005).

The individual firm, or set of competing firms in one market, are limited in their ability to sense, understand and respond to their environment compared to a network of firms spanning different markets and industries. The underlying principle is Ashby's Law of Requisite Variety. The variety of an individual firm's responses is limited by the capacity of its management to sense, understand and respond to its environment, a network of firms has a potentially far greater variety of responses, which is limited only by the alternative configurations of relations and interactions that can develop within and between them.

There are many examples of the increasing attention being given to the role and importance of business relations and networks in driving performance, competitiveness, innovation and adaptability. The paper by Hakansson in this special issue includes an account of some of the research and thinking of the IMP group about the nature and importance of relations and networks and I will not repeat them here. Two examples will suffice. One is the recognition that innovation takes place more between firms and other organisations, rather than within them.[44] The relevant interactions here are not limited to those associated with a particular market but cut across markets both horizontally and vertically and across technologies, regions and nations and include interactions with suppliers, customers, competitors and complementors. Interaction and feedback effects with the natural environment are becoming an increasing focus of attention, setting ultimate limits to the long term evolution and sustainability of economic systems.[45]

---

[44] See for example Henry Chesborough OPEN INNOVATION (2003), John Hagel, and John Seeley Brown Productive Friction HARVARD BUSINESS REVIEW February (2005), Subroto Roy, K. Sivakumar, and Ian F. Wilkinson Innovation Generation in Supply Chain Relationships – A Conceptual Model and Research Propositions J. OF THE ACADEMY OF MARKETING SCIENCE 32 (2004).

[45] Jared Diamond COLLAPSE: HOW SOCIETIES CHOOSE TO FAIL OR SUCCEED (2004).

The second example is the concept of the extended phenotype or extended enterprise and soft assembled strategies.[46] Genes, as Dawkins, has pointed out, are not limited in their expression to the body and behavior of the phenotype it happens to be embedded in.[47] They are also capable of affecting the development and behavior of other animals and plants and this has implications for survival and reproduction and hence affects the evolutionary process. He refers to the effects of genes from one animal or plant on linked plants and animals as the extended phenotype. In the same way firms and people make use of the resources and innate characteristics of their local environment and network as an extension of their senses, memory, resources and mind, such that the boundary of the self or firm is negotiable and flexible.[48]

Unfortunately most business disciplines focus attention primarily on the management of the individual firm and what drives its efficiency, competitiveness and performance. This is understandable as firms employ most of the graduates from our business programs, help fund these programs and managers are the main buyers of consulting services. But the types of environment in which firms now operate tend to generate problems and opportunities that are beyond the ability of firms to sense, comprehend or respond to independently. Such environments are variously described as turbulent, complex or hypercompetitive. In such conditions the relevant units of analysis are not individual firms but networks of interrelated and interdependent firms and organisations spanning industries, markets, technologies and nations that create major strategic problems and opportunities for firms and policy makers alike.

---

[46] Wilkinson and Young *supra* (2005).

[47] Richard Dawkins THE EXTENDED PHENOTYPE (1983).

[48] David Lane and Robert Maxfield Strategy under Complexity: Fostering Generative Relationships LONG RANGE PLANNING 29 (1996).

Complexity theory suggests that antitrust policy needs to focus attention more on the nature and role of these networks and on the different types of interaction and feedback mechanisms operating if it is to participate meaningfully in the business system's development and evolution. But the relevant interactions are not under the direct control or influence of government policy and governments have to learn how to recognise qualitatively different types of evolutionary pathways and tipping points, so as to help steer business away from pathological attractors. This, of course, much easier to say than do but research in complexity provides some guidelines.

Hakansson, in his paper in this special issue, explains how a focus on interactions and networks changes our view of the way firms and markets behave. Markets are seen as networks of exchange relations developed between firms in and across markets that shape the way a particular market behaves and evolves. The relations and networks that exist as a result of history constrain and direct the behavior of those involved and, at the same time, the experience and outcomes of behaviour reproduce, strengthen, weaken or change these relations and networks. In this way the ongoing patterns of action and interaction over time create the organisations and networks comprising a business system, as is further elaborated in the next section.

C    *Dissipative Structures*

So far I have taken the existence of economic actors, usually firms, as given and focused on the role and importance of the connections between them as driving performance, innovation and adaptation. But complexity research directs our attention to the very nature and existence of the actors in complex adaptive systems. It makes the process primary, the continual flow of action and interaction, and economic actors like firms are produced out of this ongoing flux. The come into existence as recognizable, reproduced patterns of action and interaction among people and objects over time, or what are called dissipative structures.

To illustrate we may use the analogy of a river. The continuous flow of water downstream in a river results in the formation of local patterns of repeated behaviour, such as eddies or whirlpools. These are local structures that are reproduced over time through an ever changing stream of water molecules following the same patterns of behaviour.  The eddies and whirlpools are macro structures that arise, in a self organizing manner, from the ongoing local interactions taking place among an ever-changing stream of water molecules in a river bed. Over time, as conditions change, due to increased or decreased water flow, erosion and local environmental impacts, the pattern and location of eddies and whirlpools changes. Eddies, whirlpools and business firms and networks are examples of what are called dissipative structures, a concept developed by Ilya Prigogene, which won him a Nobel Prize.   Dissipative structures are what complexity is about.

In this view <u>*non-change*</u> rather than change becomes problematic.[49]  Why do firms and networks which embody particular characteristic patterns of behaviour persist? How are these patterns reproduced over time amidst the constant flux of actions and interactions taking place? This approach offers, I believe, additional insights for guiding antitrust policies.

Antitrust policy focuses on the structure and conduct of firms in markets as a means of improving overall economic performance.  The underlying assumption is that firms organize the activities taking place in markets and the number of firms in a market and the way they individually behave matters.  After all, isn't it firms who make decisions about how and when to act and interact?  An alternative perspective focuses attention on the ongoing processes taking place, the actions and interactions, and how these create firms rather than the way firms create the patterns of actions and interactions.[50] The business of antitrust is to help shape the kinds of interaction and feedbacks that lead to the right kinds of firms and networks evolving, rather than controlling the firms and networks that happen to exist at a particular time and place in order to produce the right kinds of interaction and feedbacks.

---

[49] Andrew Abbott TIME MATTERS: ON THEORY AND METHOD (2001).

D    *Networks Matter*

The foregoing discussion highlights the role and importance of networks in business systems. Antitrust theory can draw on research on the nature and evolution of networks to inform its understanding of business networks and how government policy could play a productive role in business evolution. Our understanding of network structures and their effects has improved significantly in recent years as a result of the study of complexity.[51] Neo-classical economic theory of perfect markets assumes a random mixing of suppliers and customers resulting in a random network of trades. If extended to other vertically linked or complementary markets a larger random network configuration would be the result. Business networks would comprise fleeting interactions among a set of equal sized small suppliers and customers and this model seems to be the basis of much antitrust policy around the world. Distortions or failures of perfect markets in the form of monopolies and imperfect competition, whether they arise for natural or contrived reasons, are a central focus of antitrust cases. It is generally assumed that such failures are not in the consumers' or societies' interests and they need to be controlled or reconfigured. Unfortunately research shows that networks of people and firms (as well as other entities) do not develop into such random network structures.

---

[50] Ian F. Wilkinson Toward a Theory of Structural Change and Evolution in Marketing Channels JOURNAL OF MACROMARKETING (1990).

[51]  See for example Ball *supra* (2005) and Albert-László Barabási LINKED: THE NEW SCIENCE OF NETWORKS(2002).

Depending on the patterns of interaction taking place and the way they are interconnected over time and place, different types of trading networks and associated patterns of exchange relations can emerge which have characteristic properties. These properties are a source of opportunities and threats. The focus for antitrust theory and policy is to understand the processes of network development and evolution and to identify ways of intervening that influence these processes in productive ways, rather than simply regulating firms occupying particular network positions that are the outcomes of these processes. The outcomes, in a sense, are inevitable; it is the processes of business development and evolution that require more attention than the structure it produces. The structure reflects the processes at work and only temporary reconfigurations are possible if the underlying processes remain unchanged.

What are the types of networks that can arise? Which ones should we be more concerned about? And what types of processes produce different kinds? These are all big questions and I do not believe we have firm answers yet but lots of intriguing work is going on. As Barabarasi, one of the leaders in this field, has commented, there is a "zoo of network types out there."[52] For our purposes three non-random types deserve our attention - structured, clustered or small world, and hub or scale free networks. Structured networks are those following a fixed regular pattern of connection, such as a lattice or grid, and usually reflect a network deliberately designed for some purpose. Such networks generally do not grow naturally as there usually is no designer or network controller. Instead, network structures emerge in a self organizing way from the patterns of interactions taking place and the way they develop over time.

---

[52] Cited in Ball *supra* (2005).

Small world networks have received much attention owing to the research of Strogatz and Watts.[53] Small world networks are those in which each node is only a few steps or links away from any other node, even though the network is not richly interconnected. Most links are clustered together around those nearby in some way e.g. space, behaviour, beliefs, and these local networks tend to be highly interlinked or clustered. People who know you well probably know each other; firms compete and trade more with those serving the same customers or using similar suppliers and inputs. But some links are with other nodes that are not part of the local network and which may be "far away" in some sense. These long reach links have the effect of reducing significantly the average distance in links between nodes in a network. The importance of these links in business and markets is reflected in Ronald Burt's work on the importance of structural holes[54] and Mark Granovetter's research concerning the strength of weak ties.[55] They show how long reach links tend to be weak links but are an important source of new information and ideas. This is because strongly interconnected people or firms are more likely to know what each other knows, whereas long reach links may bridge structural holes in networks, acting as bridges between otherwise non connected networks or parts of a network. For example, the nature and role of entrepreneurs can be explained in part by the way people or firms are positioned in communication networks such that they gain access to potentially productive assortments of knowledge and ideas that others do not. By bridging structural holes in the network they are able to gain strategic advantages in terms of opportunity discovery and exploitation. Unfortunately, it is not yet possible to identify all the structural holes that might be occupied productively, or to estimate how many of them have been occupied – but the internet and the death of distance is having an effect I am sure.

---

[53] Duncan Watts SMALL WORLDS (1999).

[54] Ronald Burt STRUCTURAL HOLES: THE SOCIAL STRUCTURE OF COMPETITION (1992).

[55] Mark Granovetter and Roland Soong Threshold Models of Interpersonal Effects in Consumer Demand J. OF EC BEHAVIOUR AND ORG 7 (1986).

The third form of network structure is the hub or scale free network, which can be confused with a small world network because it does have small world properties – nodes are on average not far from each other in terms of number of links. But scale free networks have a distinctive distribution of links per node called a power law, which is quite different to the normal or Gaussian distribution we are familiar with from courses in statistics. If the frequency distribution of links per node followed a normal distribution, we would expect most nodes to be around the middle of the distribution with a tail on either side. The central tendency or mean reflects the scale of the distribution or where it is located on a scale of number of links per node. A power law has no central tendency or characteristic scale, instead most nodes have very few links and the frequency of nodes with x links falls off rapidly as x increases. On a log log scale, showing the cumulative frequency of nodes by number of links, the curve is a straight line. This means that some links have a very large number of links, while most do not. These are the hubs in the network. Such network patterns have been found in many situations including earthquakes, ecological systems, cities, friendship networks, sexual partners, movie actors, co-authorships, patent links and trading networks.

Iansiti and Levien in their analysis of business ecologies refer to hubs in this type of network structure as keystones, with Microsoft being an important example.[56] This has led to some controversy over whether hubs are a good or bad thing in economic terms, especially with the recent Antitrust Case concerning Microsoft. Such a network position seems obviously good for Microsoft but is it good for the US and the world economy to have such hubs? Can and should Microsoft's behaviour and the structure of the networks in which it is involved be changed or controlled in some way and who and how should this be done? The increasing globalization of the world's economies makes business systems and networks span the world rather than being weakly interconnected national business systems. This makes the problem of developing and implementing antitrust policies even more difficult, as they involve policy coordination across nations.

---

[56] Marco Iansiti and Roy Levien THE KEYSTONE ADVANTAGE (2004).

Research on the nature and formation of networks offers some guidance. First, such scale free networks seem to be characteristic of any living system and reflect the way networks grow naturally. The underlying rule for development may be summarized in terms of the "the rich get richer." As a network grows new nodes are more likely to form links with, be attracted to, or find, nodes that are themselves already well connected. The probability of a new node linking with another node depends on how many links the other node already has. Thus people and firms are more likely to form links with people and firms who already have more links with other people and firms. Various psychological, social and economic mechanisms lead to this type of behaviour, which represent forms of positive feedback effects. In the case of Microsoft, the network economies involved in using a computer operating system that many already use is the main positive feedback mechanism.

Trading or exchange networks are likely to form scale free networks because they grow by similar processes to those described. In markets the pattern of market shares among brands and firms follows a power law type distribution, reflecting the frequency of purchase and amount bought by customers over a period. The trading relations firms have with their suppliers and customers is usually characterized in terms of the 80:20 rule whereby 20% say of its customers account for 80% of its business. The number of trading relations firms have with others firms might also follow some kind of power law distribution although data on this is limited. In economic theory the development of scale-free networks is reflected in models of markets with strong network externalities, which is a special type of rich get richer development rule. These markets result in outcomes in which the winner takes most of the market - a scale free network with hubs.[57]

Hence we should expect power law distributions and hubs to arise. It is not the property of the firm or person occupying the hub that causes it to become a hub, it is an outcome of the way networks grow and evolve. How should we deal with them and what are the benefits and dangers? Are they another form of natural monopoly that requires government regulation, as is sometimes argued? Hubs are a type of network role or position that emerges naturally in any living business or economic system. What matters is how easy it is for rivals to take over hub positions from others if they can offer better value. This is the main worry for antitrust policy makers. The degree of competition for hub roles, by definition, is not indicated by the presence of many equally strong rival hub positions in a network as this would not be a scale free network. Hub competition is reflected in how contestable the market for hub roles are, which refers to the ease or difficulty with which other firms are able to move into hub roles.[58] This includes existing minor hubs in the network, which in the case of Microsoft would include alternative operating systems such as Apple, Linux and UNIX, as well as organizations occupying hub or keystone positions in related networks, such as Google, Amazon, PC or internet connection providers.

Governments can move into hub roles by taking over and controlling hub firms in the public interest. However, it is not clear that government managed hubs are desirable in evolutionary terms. If firms occupying hub roles are subject to government takeover and the free release of core technology, they have incentives to disguise their "hubness" by promoting other hubs. This leads to a more visibly competitive hub market but how efficient and effective would it be and how would this affect the evolution of the network over time? Another policy option is to break up hub firms, but if scale free networks are a natural result of interactions, breaking up Microsoft will just lead to the emergence of another operating system assuming the same powerful position. Is this the way to go or would it damage evolvability?

---

[57] Nicholas Economides, Nicholas The Microsoft Antitrust Case J. OF INDUSTRY COMMERCE AND TRADE 1 (2001).

[58] The theory of contestable markets is described in William J. Baumol, John C. Panzar and Robert D. Willig CONTESTABLE MARKETS AND THE THEORY OF INDUSTRY STRUCTURE (1982).

Can hubs become too powerful and control and direct the future evolution of the network in their own interests and against the larger good, or are there natural constraints that limit the power and reach of economic hubs or keystones like Microsoft. I have already mentioned the contestability of hub positions as one limiting factor. Others are suggested in research by Barabarasi.[59] For example, there are limits to how many movies one actor can be in, how many friends an individual can have. Firm size, technology and geographic constraints affect how many customers or suppliers a firm can handle. In addition, firms and people are heterogeneous and there are forces of repulsion as well as attraction in networks, such that at times the rich may get poorer, at least for some in the network. For example, some people have strong negative feelings about using Microsoft and are committed to other operating systems such as Apple or UNIX.

Microsoft's role is similar to an organisation being in control of the development and evolution of the English language. You don't have to use English but many people are using it because so many people already use it in business, science and social life around the world. We might argue about whether we all speak the same English but let's not get into that. Languages evolve as new words are added, their meaning changes and the rules of grammar alter. All of us are free to use words to say, write and think what we want in English but there are some things that are more easily said and thought in other languages, so I believe. No one controls the English language and regulates the core rules of grammar, spelling and meaning of words, although dictionaries and rules of good English are produced.  Indeed, its living nature comes from the way these standards change over time and how innovations in language arise and spread.  Attempts to control a language tend to fail or be counterproductive , as has happened for example with the French language.  Despite the attempts to control it the French language has evolved and in ways not necessarily desired by the French government.  Does this mean Microsoft is unable to control the evolution of operating systems and the way they are used?  From my limited understanding of the technicalities involved Microsoft seems potentially more powerful than the French government is in controlling French. It power comes from the core of its operating system, which is only made available to developers in machine code form such that it cannot be altered by anyone other than Microsoft.  If this is so evolvability is constrained.  How serious this is and how difficult it is to work around this constraint is hard to say.  Operating systems like languages evolve and people and firms can switch operating systems and languages when the value of an alternative becomes greater and switching costs do not wipe out the benefits.

---

[59] Barabarasi *supra* (2002).

There is no simple answer to all these questions and problems but complexity theory and its associated methodologies help us address them more clearly.  In the remainder of this article I will consider the potentially good and bad characteristics of a scale free business networks, their evolution and evolvability and the role antitrust policy could play.

IV     Nature, Role and Regulation of Scale Free Business Networks

Scale free networks are resilient because random failures of parts of the network do not affect its connectivity and functioning very much.  However, if hubs can be identified such networks are vulnerable to attack.  This implies that we need to be worried about the failure of a Microsoft, as this could lead to a rapid contagion and damage throughout the networks of organizations directly and indirectly linked to it.  As a keystone Microsoft is equally good at enabling value creation in others and in rapidly spreading damage by inhibiting and biasing the efforts of others.  It is the latter possibility that lies at the heart of much discussion about regulating Microsoft. Microsoft occupies a hub position in business networks that gives it the power to do good or evil and most economic theory assumes that monopolists like this will use their power in their own interests and against the interests of the many.

We need to distinguish between the effects of regulating the behaviour or breaking up <u>existing</u> hubs on the performance and behaviour of others in the network and effects on the evolution and evolvability of the network. For example, will regulating a hub tend to ensure its continuation as a hub, preventing other forms of network evolution? Will breaking it up lead to the emergence of another hub with similar characteristics, which then has to prosecuted all over again, or will it lead to some process of network evolution that will be better or worse? In order to answer these questions we need some idea of how networks evolve over time. After all, there used to be business networks before Microsoft.

Business networks evolve through a process of reconfiguring the links among actors (business, government, lawyers, and educators), activities, resources and schemas. Incremental changes are taking place all the time through the ongoing flow of activities taking place. Resources are being used and created, strengthened and weakened, learning and knowledge development and diffusion is taking place. Radical changes involve significant and novel types of reconfiguration. How does a scale free network affect the evolutionary process? Innovations involve generating new ideas and which come from recombining existing ideas – they are not endogenously determined manna from heaven.[60] Ideas get recombined through the research effort of individual actors and through the communication and interaction-taking place among and within actors.

---

[60] Richard R. Nelson and Paul M. Romer Science, Economic Growth and Public Policy in B. Smith and C. Barfield eds. TECHNOLOGY R&D AND THE ECONOMY (1996).

Scale free networks help preserve ideas, both good and bad, because ideas are hard to eliminate completely. This is shown by research on the spread of viruses in human and internet environments. Although programs targeted at hubs can significantly reduce their presence and spread in the network, they remain in parts of the network ready to spring into action when network conditions permit. Scale free networks also facilitate the recombining of ideas because of their small world character. This helps ideas to be found and for ideas to find productive homes through communication networks. No matter where ideas are located, they are not very far from others in the network. This recombining is limited by the extent to which ideas are easily found, communicated and used by others and by the way knowledge is distributed throughout the network. Some ideas are locked away by firms, protected by patents or not easily communicated or understood because they are tacit, sticky and embedded in relations and routines.[61].

Hubs play an important part in facilitating the spread of ideas and therefore influence innovation through impacts on the opportunity discovery and exploitation process. As this is central to evolution and the evolvability of the network this subject should play an important role in antitrust policy deliberations. This suggests that potential targets for antitrust policy are the various types of actual and potential transmission biases that could adversely affect evolvability. Some of these already form part of antitrust policy because they are related to issues of static efficiency and price competition. For example, misleading and deceptive practices result in biased transmission and affect the numbers and types of competitors which in turn limit the cultural variants for the evolutionary processes to work on.

Other ways in which potentially productive recombination of ideas can be inhibited include: entry barriers to new types of firms; obstacles to new types of relations and interactions forming among firms in and among different markets, industries and nations; the ability of workers to set up or move to new firms or nations in order to exploit innovations that their existing firms or nations are unwilling or unable to exploit; the limited exposure of some industries and parts of networks to international cooperative and competitive interactions; and the inability and unwillingness of firms to share information and ideas with other firms for fear of losing control of the ideas and/or damaging their own prospects compared to counterparts. These are perennial problems that shape the kinds of business networks that arise, the firms that eventually become hubs, how these change over time and how efficient a given network is at any given time.

---

[61] J.L. Badaraco THE KNOWLEDGE LINK: HOW FIRMS COMPETE THROUGH STRATEGIC ALLIANCES (1990).

The central issue is not to freeze business networks and antitrust policy in terms of a particular structure, pattern of conduct and/or regulation, but to ensure many natural experiments can and do take place among and within firms and in different policy domains, in order to give evolution more to work with. In this way the requisite variety of network and business systems is preserved and many potential evolutionary pathways are opened up rather than closed off. Freezing a network or part of it is like killing off part of the living system, reducing it to be part of the environment rather than part of the living system and evolutionary process.

*A     Agent Based Models of Markets Matter*

Tools to help us map and model the structure and evolution of business networks are now available to facilitate policy development and case analysis.[62] Of particular relevance are agent based models of economic systems which offer ways of developing and testing our theories of business structure, conduct and evolution including the impact of various types of antitrust policies.

---

[62] See for example Walter W. Powell, Douglas R. White, Kenneth W. Koput, and Jason Owen-Smith Network Dynamics and Field Evolution: The Growth of Interorganizational Collaboration in the Life Sciences AMERICAN JOURNAL OF SOCIOLOGY 110 (2005).

We cannot conduct experiments about the impact of different types of antitrust policies on actual economic systems. It would be politically and managerially impossible to implement, extremely costly, be likely to damage whole industries and economic sectors, and take a long time. Instead, we need to construct evolutionary models of business networks that can be used to conduct such experiments, that can be used help us map out the potential range of outcomes implicit in particular network structures and processes and contexts, and that can help us understand what antitrust policy can and cannot do. These models will include antitrust policy development and enforcement as part of the model, not standing outside, godlike, and merely imposing new rules as the occasion arises in the model. Developments in computing systems and agent based modeling methods enable us to represent key aspects of complex adaptive systems, such as economic and business systems, and the ways they develop and evolve that hitherto have not been possible. These models allow managers and policy makers to develop and hone their experience, sensitivity and understanding of the complex systems in which they operate by providing realistic "flight simulator" type interfaces for them work with. Systematic experiments can be conducted in these artificial worlds to learn about the range of outcomes possible under different conditions, including different types of intervention mechanisms, in ways analogous to biological and chemical experiments.[63] An example of these kinds of models is an agent based model of the wholesale electricity markets in the US being developed by Leigh Tesfatsion and her colleagues, which is designed to test the economic reliability of the Wholesale Market Power Platform proposed by the US Federal Energy Regulatory Commission in 2003.[64] Such models will focus the attention of managers, policy makers and lawyers on key issues and assumptions that affect the development and evolution of the relevant business networks and their outcomes. In this way we can move beyond the existing narrow focus on static efficiency and price competition to considerations of dynamic efficiency, value and evolvability.

---

[63] For a more extended discussion of these types of models see Leigh Tesfatsion, Agent-based Computational Economics: A Constructive Approach to Economic Theory in Tesfatsion and Judd (eds.), HANDBOOK OF COMPUTATIONAL ECONOMICS, VOLUME 2: AGENT-BASED COMPUTATIONAL ECONOMICS (2005).

V    Conclusions

I have argued that the main case for antitrust policy should be extended beyond its current focus on economizing (including both production and transaction costs), including market power as a key filter for identifying suspect cases. The economy is a complex adaptive system and research on such systems focuses attention on their dynamics, evolution and network structure. I suggest "evolvability" as an additional main case consideration which leads to a focus on networks costs and benefits  Transaction costs (and benefits) focus on individual transactions and relations in isolation. Network costs (and benefits) focus attention on the *connections between transactions and relations over time and place*, including how business arrangements at one time limit or enable arrangements in the future. Such considerations can be included in the rules of antitrust and be included in the process of antitrust case analysis and decision making.

Incorporating such ideas into antitrust policy is not without its difficulties. In particular it requires that they are administratively and politically feasible. This of course has to do with the evolvability of the law, policymakers and lawyers, as well as business and law schools that are involved in the production and diffusion of relevant business memes! Antitrust policy apparatus, business schools and law schools do not stand outside business systems but are relevant and important actors in business ecosystems. They affect and are affected by business practice and ideas about business practice. Furthermore, in an increasingly globalized world business system, issues of evolvability and economizing are not confined to one nation but to interrelated networks of nations and their antitrust policies. US antitrust policy and decisions impact other countries business systems and vice versa, which leads us into a much larger evolutionary arena.

---

[64] D. Koesrindartoto, J. Sun and Leigh Tesfatsion An Agent-Based Computational Laboratory for Testing the Economic Reliability of Wholesale Market Designs INTN'L CONF ON COMPUTING IN ECON AND FINANCE WASHINGTON, June (2005).

A way of facilitating policy development and case analysis is through the use of agent based models of relevant business networks. These are capable of extending our understanding, analysis and theory in the area of business networks and their evolution and are a means of focusing policy debate and disputes in productive ways. Agent based models are more than a technique; they are a modeling philosophy and a way of developing, analyzing and comparing theory and practice not possible in other ways.

By introducing an evolutionary perspective, antitrust policies are seen in a new and helpful light and attention is directed to potential policy targets and mechanisms that may not otherwise be considered. In order to do this we need to make use of methods and theories of business and economics drawn from complexity and network research. We are at an exciting stage of development of these ideas, theories and methods and need to draw from them and contribute to them.

**Figure 1 The Evolutionary Process***

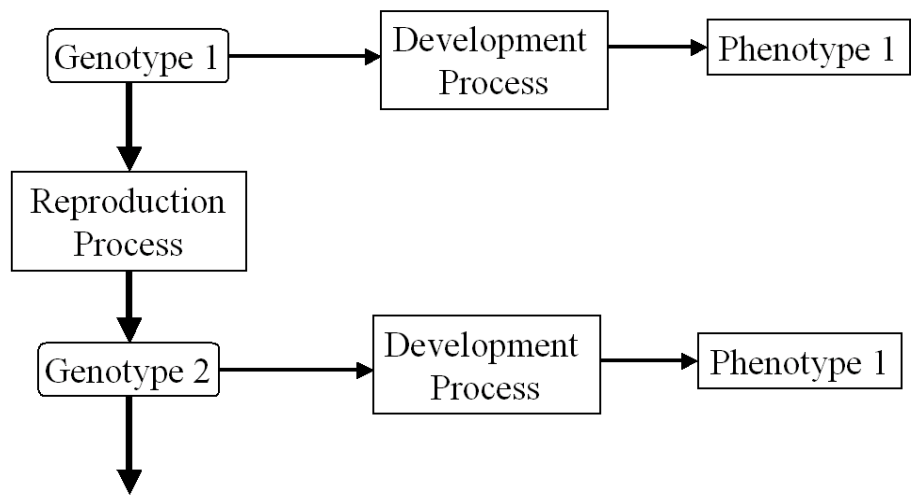

(Adapted from R. Dawkins 1998 p 202)

**Table 1 Components of Evolution Processes in Biology and Business**

| Dimension | Biology | Business |
|---|---|---|
| **Replication Unit** | Genes | Memes |
| **Transmission Unit** | Genotypes | Memotypes |
| **Phenotype** | Flora and Fauna | Firms, Households & Organisations |
| **Transmission Process** | Sex, Division | Social and Economic Learning and Copying |
| **Adapting Unit** | Extended Phenotype | Network |
| **Variation** | Mutation & Recombination | Innovation, Error & Recombination |